\documentclass[fleqn,usenatbib]{mnras}

\usepackage{newtxtext,newtxmath}
\usepackage{hyperref}
\usepackage{academicons}
\usepackage{xcolor}

\usepackage[T1]{fontenc}
\usepackage{enumerate}
\DeclareRobustCommand{\VAN}[3]{#2}
\let\VANthebibliography\thebibliography
\def\thebibliography{\DeclareRobustCommand{\VAN}[3]{##3}\VANthebibliography}
\newcommand{\orcid}[1]{\href{https://orcid.org/#1}{\textcolor[HTML]{A6CE39}{\aiOrcid}}}

\usepackage{graphicx}	
\usepackage{amsmath}	
\usepackage{comment}

\newcommand{\maxi}{MAXI~J1957$+$032}
\newcommand{\nicer}{NICER}

\newcommand{\chandra}{Chandra}
\newcommand{\swift}{Swift}
\newcommand{\swiftxrt}{Swift/XRT}
\newcommand{\swiftuvot}{Swift/UVOT}
\newcommand{\inte}{INTEGRAL}
\newcommand{\maxigsc}{MAXI/GSC}
\newcommand{\maxit}{MAXI}

\title[New AMXP \maxi{}]{\maxi{}: a new accreting millisecond X-ray pulsar in an ultra-compact binary}

\author[A. Sanna et al.]{
A.~Sanna$^{1}$\thanks{E-mail: andrea.sanna@dsf.unica.it} \href{https://orcid.org/0000-0002-0118-2649}{\includegraphics[scale=0.08]{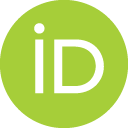}},P.~Bult$^{2,3}$\href{https://orcid.org/0000-0002-7252-0991}{\includegraphics[scale=0.08]{ORCIDid}}, M.~Ng$^{4}$\href{https://orcid.org/0000-0002-0940-6563}{\includegraphics[scale=0.08]{ORCIDid}}, P.~S.~Ray$^{5}$\href{https://orcid.org/0000-0002-5297-5278}{\includegraphics[scale=0.08]{ORCIDid}}, G.~K.~Jaisawal$^{6}$\href{https://orcid.org/0000-0002-6789-2723}{\includegraphics[scale=0.08]{ORCIDid}}, L.~Burderi$^{1}$\href{https://orcid.org/0000-0001-5458-891X}{\includegraphics[scale=0.08]{ORCIDid}},  T.~Di Salvo$^{7}$\href{https://orcid.org/0000-0002-3220-6375}{\includegraphics[scale=0.08]{ORCIDid}}, \newauthor A.~ Riggio$^{1,8}$\href{https://orcid.org/0000-0002-6145-9224}{\includegraphics[scale=0.08]{ORCIDid}}, D.~ Altamirano$^{9}$\href{https://orcid.org/0000-0002-3422-0074}{\includegraphics[scale=0.08]{ORCIDid}}, T.~E.~Strohmayer$^{3}$\href{https://orcid.org/0000-0001-7681-5845}{\includegraphics[scale=0.08]{ORCIDid}}, A.~Manca$^{1}$, K.~C.~Gendreau$^{3}$, D.~Chakrabarty$^{4}$\href{https://orcid.org/0000-0001-8804-8946}{\includegraphics[scale=0.08]{ORCIDid}},\newauthor W.~Iwakiri$^{10}$\href{https://orcid.org/0000-0002-0207-9010}{\includegraphics[scale=0.08]{ORCIDid}}. R.~Iaria$^{7}$\href{https://orcid.org/0000-0003-2882-0927}{\includegraphics[scale=0.08]{ORCIDid}}
\\
$^{1}$Dipartimento di Fisica, Universit\`a degli Studi di Cagliari, SP Monserrato-Sestu km 0.7, 09042 Monserrato, Italy\\
$^{2}$Department of Astronomy, University of Maryland, College Park, MD 20742, USA\\
$^{3}$Astrophysics Science Division, NASA Goddard Space Flight Center, Greenbelt, MD 20771, USA\\
$^{4}$MIT Kavli Institute for Astrophysics and Space Research, Massachusetts Institute of Technology, Cambridge, MA 02139, USA\\
$^{5}$Space Science Division, Naval Research Laboratory, Washington, DC 20375-5352, USA\\
$^{6}$National Space Institute, Technical University of Denmark, Elektrovej 327-328, 2800 Lyngby, Denmark\\
$^{7}$Universit\`a degli Studi di Palermo, Dipartimento di Fisica e Chimica, via Archirafi 36, 90123 Palermo, Italy\\
$^{8}$INAF/IASF Palermo, via Ugo La Malfa 153, I-90146 - Palermo, Italy\\
$^{9}$School of Physics and Astronomy, University of Southampton, Southampton, Hampshire SO17 1BJ, UK\\
$^{10}$Department of Physics, Faculty of Science and Engineering, Chuo University, 1-13-27 Kasuga, Bunkyo-ku, Tokyo 112-8551, Japan\\	
}

\date{Accepted 2022 August 11. Received 2022 August 5; in original form 2022 July 15}

\pubyear{2022}

\begin{document}
\label{firstpage}
\pagerange{\pageref{firstpage}--\pageref{lastpage}}
\maketitle

\begin{abstract}

The detection of coherent X-ray pulsations at $\sim$314~Hz (3.2~ms) classifies \maxi{} as a fast-rotating, accreting neutron star. We present the temporal and spectral analysis performed using \nicer{} observations collected during the latest outburst of the source. Doppler modulation of the X-ray pulsation revealed the ultra-compact nature of the binary system characterised by an orbital period of $\sim1$ hour and a projected semi-major axis of 14~lt-ms. The neutron star binary mass function suggests a minimum donor mass of $1.7 \times 10^{-2}$~M$_{\odot}$, assuming a neutron star mass of 1.4~M$_\odot$ and a binary inclination angle lower than 60 degrees. This assumption is supported by the lack of eclipses or dips in the X-ray light curve of the source. 
We characterised the 0.5--10~keV energy spectrum of the source in outburst as the superposition of a relatively cold black-body-like thermal emission compatible with the emission from the neutron star surface and a  Comptonisation component with photon index consistent with a typical hard state. We did not find evidence for iron K-$\alpha$ lines or reflection components. 
\end{abstract}

\begin{keywords}
binaries:general–stars:neutron – X-rays:binaries – accretion: accretion disks
\end{keywords}



\section{Introduction}

\label{sec:introduction}
Accreting millisecond X-ray pulsars (AMXPs) are rapidly-rotating (spin frequency $>$ 30 Hz) neutron stars (NS) gravitationally bound with late-type companion stars \citep[see e.g.,][for extensive reviews]{Di-Salvo:2020va,Patruno:2021vs}. Their distinct short spin periods are a direct consequence of prolonged mass transfer phases in which the companion star loses matter via Roche-lobe overflow, subsequently accreted onto the NS \citep[{\it recycling scenario};][]{Alpar82}. The sample currently includes 24 sources, a third of which are characterised by an orbital period shorter than 80 minutes (also known as \emph{ultra-compact} binaries). Short orbital periods suggest small low-mass companion stars, consistent with donor masses on average <0.2~M$_{\odot}$.

\maxi{} was observed for the first time by \maxit{} \citep{Negoro:2015vw} and \inte{} \citep{Cherepashchuk:2015vo} in May 2015. At odds with standard low-mass X-ray binaries (LMXBs) \maxi{} exhibited four short ($<5$ days) faint outbursts between its discovery and October 2016 \citep{Sugimoto:2015wh, Tanaka:2016we}. Optical observations of the system during its 2016 outburst revealed emission compatible with an irradiated X-ray disc in a LMXB, suggesting similarities with AMXPs \citep{Mata-Sanchez:2017vl}. An optical counterpart (late-K/early-M dwarf star) has been identified during the X-ray quiescence phase, setting a constraint on the source distance of the order of $5\pm2$~kpc \citep{Ravi:2017tl}. However, the outburst properties of \maxi{} let the author to suggest that the observed counterpart is likely not the mass donor. Instead, it is suggested to be a possible triple system, with the main-sequence counterpart likely being in a wide orbit around a compact interacting binary. Moreover, the spectral evolution investigated by combining the \swift{} observations collected during the four outbursts suggested that, for a distance of the order of 4~kpc, a neutron star (NS) might be hosted in the binary system \citep{Beri:2019va}. 

On June 18, 2022, \maxigsc{} detected X-ray activity in the direction of \maxi{} \citep{Negoro:2022tb}. \nicer{} quickly started monitoring the new outburst, discovering coherent X-ray pulsations at $\sim$314 Hz and revealing the nature of the accreting compact object \citep{Ng:2022uc}. A preliminary orbital solution obtained from continued \nicer{} observations suggests that \maxi{} is an \emph{ultra-compact} binary with an orbital period of $\sim1$ hour \citep{Bult:2022tv}.

\swiftxrt{} observations on June 20, 2022, suggested spectral properties consistent with the 2016 outburst of the source \citep[see, e.g.][]{Beri:2019va}. Moreover, \swiftuvot{} detected the UV counterpart of the source, with a magnitude UVW2 = $20.37\pm0.08$ \citep{Beri:2022wp}. On June 21 \maxi{} was not detected in radio by the MeerKAT observatory, with an upper limit of 48 $\mu$Jy  \citep{van-den-Eijnden:2022vf}. Optical observations suggested the presence of short timescale variability during the outburst \citep{Baglio:2022ut}, and a significant optical counterpart  during the 50 days before the outburst, with the first optical brightening almost nine hours earlier to the \maxigsc{} trigger \citep{Wang:2022us}. On June 23, \swift{} revealed no significant X-ray activity at the source location, suggesting that \maxi{} may have entered the quiescence phase \citep{Chandra:2022wx}.

Here, we report on the discovery of millisecond X-ray pulsations from the X-ray transient \maxi{} and its spectral properties from the \nicer{} observations collected during its latest outburst.

\section{Observations and data reduction}

\maxi{} was observed by \nicer{} X-ray telescope on the International Space Station \citep{Gendreau:2016tk} from June 19, 2022 to June 24 (ObsIDs 5202840101-6) for a total exposure time of $\sim 22.5$~ks after standard filtering. We also included a short data segment ($\sim40$ seconds at MJD $\sim$59740.1) collected during a raster scan performed to better constrain the source position. We retained events in the 0.5--10~keV energy band by processing the observations with HEASoft version 6.30.1 and the \nicer{} software \textsc{NICERDAS} version 9.0 (2022-01-17$\_$V009) with standard screening criteria. 
We extracted source and background spectra in the 0.5--10~keV energy range using the \textsc{nibackgen3C50} tool \citep{Remillard:2022tt}, and we generated response matrices using the \textsc{nicerrmf} and \textsc{nicerarf} tools. The analysis was performed with Xspec 12.12.1 \citep{Arnaud96} after applying an optimal binning, which guarantees at least 25 counts per energy bin. 

The top panel of Figure~\ref{fig:timing} shows the background-subtracted light curve of the outburst monitored by \nicer{} (black points). Each point represents the 16s average count rate. ObsIDs 5202840105-6 are not shown since the source count rate is compatible with the background. The source count rate at the peak of the observed outburst is $\sim 160$ cts~s$^{-1}$, which exponentially decreases to the quiescence level in almost four days. No Type-I thermonuclear X-ray bursts were observed during the observations.
Finally, we applied barycentric corrections to the photon arrival times utilising the \textsc{barycorr} tool adopting JPL DE-405 Solar system ephemeris. We considered the best available source position coordinates obtained by \chandra{} during the 2015 and 2016 outbursts \citep[][]{Chakrabarty:2016vp}. 

\section{Results}
\subsection{Timing analysis}
\label{sec:timing}

\begin{figure}
\centering
\includegraphics[width=0.5\textwidth]{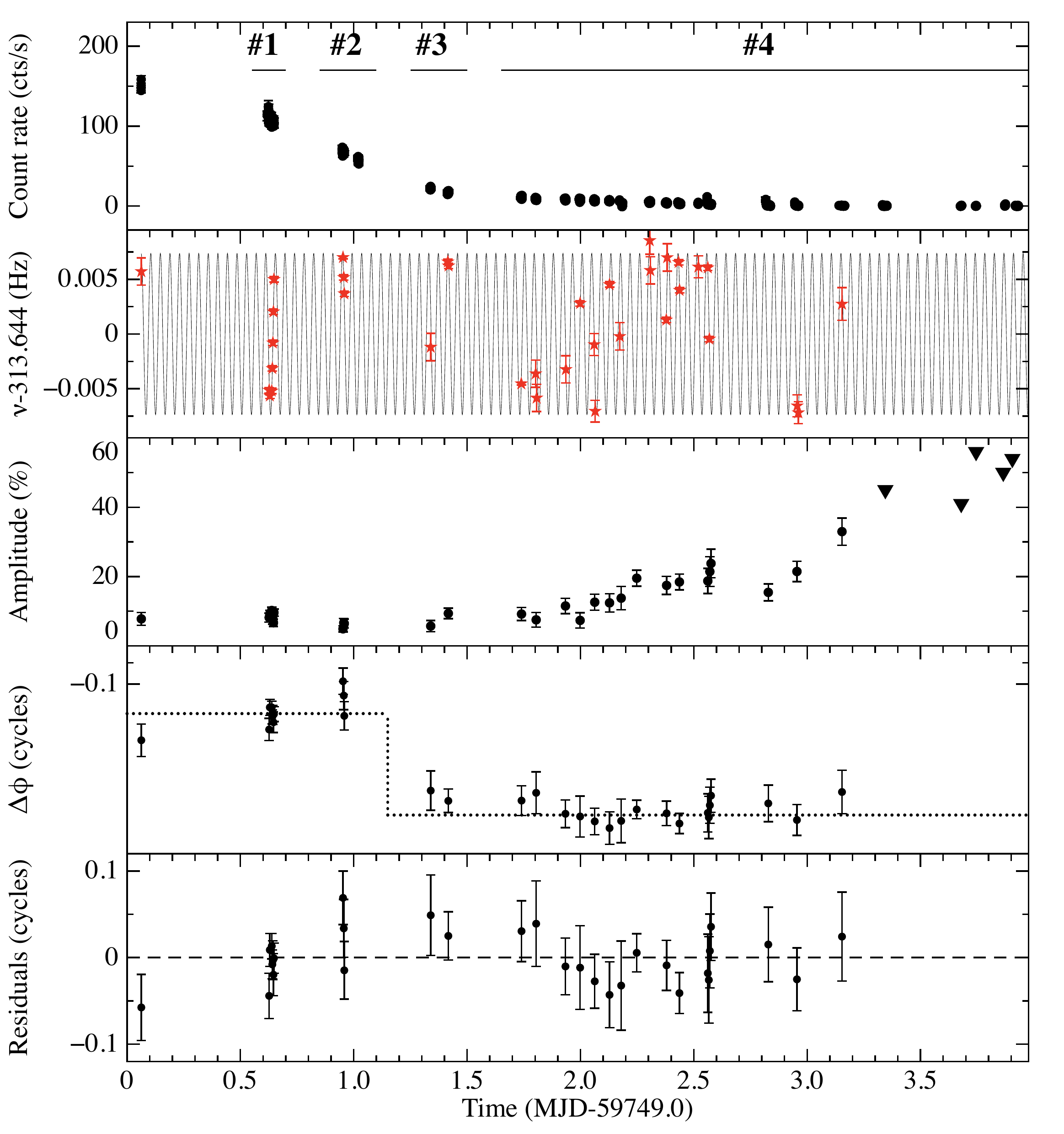}
\caption{\textit{First panel -} \nicer{} 0.5--10~keV light curve of the June 2022 outburst of \maxi{}. Filled circles represent 16s average background-subtracted count rate of the source. \textit{Second panel -} Temporal evolution of the pulsar frequency (with respect to $\nu=313.644$ Hz) estimated from \nicer{} data segments. The solid black line represents the best-fitting orbital Doppler modulation assuming a circular orbit. \textit{Third panel -} Evolution of the fractional pulse amplitude estimated in the energy range 0.5--10~keV (filled circles), and upper limits ($3\sigma$ c.l.) on the non-detection (filled triangles). \textit{Fourth panel -} Evolution of the pulse phase delays obtained by epoch-folding the \nicer{} photon arrival times corrected for the best-fitting orbital solution from the phase-coherent analysis. The dotted line serves the purpose of highlighting the phase jump observed around 59750.2 MJD. \textit{Fifth panel -} Residuals in pulse cycles with respect to the best-fitting models for the pulse phase delays. }
\label{fig:timing}
\end{figure}

Following the detection of X-ray pulsations at $\sim 314$ Hz \citep{Ng:2022uc}, we proceeded by searching for X-ray pulsations over short time intervals (between 150 and 500 seconds, depending on the statistics of the data segment) by performing epoch-folding search techniques using 8 phase bins and starting with the spin frequency value $\overline{\nu}=313.643740$ Hz \citep{Bult:2022tv}. We explored the frequency space with a $10^{-5}$ Hz frequency step for a total of 10001 steps. We detected significant X-ray pulsations in 38 of the 60 data segments. The signal frequency value and uncertainty for each data segment have been determined following the method described by \citet{Leahy1987}. The temporal evolution of the detected signal (second panel of Figure~\ref{fig:timing}, red stars) is compatible with an orbital modulation. Assuming a circular orbit, we obtained the best-fit for an orbital period of $P_\mathrm{orb}=3653.47(69)$ seconds, a projected semi-major axis of the NS orbit $x = 0.01367(35)$ lt-s, an epoch of ascending node passage $T_{ASC}=59749.63327(22)$ MJD, and a spin frequency $\overline{\nu}=313.64373(12)$ Hz (Figure~\ref{fig:timing}, second panel, solid line).   

We then proceeded with the phase-coherent timing analysis by generating pulse phase delays in the time interval between MJD 59749.625494 and MJD 59749.648643 (June 19; for an exposure time of $\sim$ 1.3~ks), where the \nicer{} count rate is higher. We folded data segments of $\sim$200 seconds into 8 phase bins at the preliminary spin frequency $\overline{\nu}$. We then modelled each pulse profile with a constant plus a sinusoidal function, and we retained profiles for which the pulse amplitude is at least three times larger than its uncertainty. We modelled the pulse phase evolution with a constant frequency combined with a circular Keplerian orbital model \citep[see][for a more detailed description of the procedure]{Sanna:2016ty}.


Due to the short baseline covered by the dataset with respect to the orbital modulation of the system and the relatively small uncertainties on the orbital period and projected semi-major axes, we only explored corrections on the spin frequency and the epoch of ascending node passage. The best-fit is obtained for $\overline{\nu}=313.6436542(61)$ Hz, and $T_{ASC}=59749.633066(17)$ MJD. We then propagated the solution to the nearest (in time) data segments verifying that the phase uncertainty remained smaller than half of a spin cycle, a condition required for the application of phase-coherent analysis. 
It is noteworthy that a similar conclusion can be reached even accounting for spin frequency derivatives $|\dot{\nu}|\leq 10^{-11}$~Hz~s$^{-1}$, orders of magnitude larger than the values observed for AMXPs \citep[see, e.g.,][]{Di-Salvo:2020va}. We generated pulse phase delays for each increased dataset and fitted them, searching for a stable timing solution, until we covered the whole outburst.         

As shown in the fourth panel of Figure~\ref{fig:timing}, the pulse phase delays from the most accurate timing solution show a phase jump of $\sim$0.2 pulse cycles around MJD 59750.2. To account for that, we included in the model a phase jump around MJD 59750.2. 
The best-fit orbital and pulsar spin parameters from the latter model are reported in Table~\ref{tab:solution}, while its associated residuals are shown in the fifth panel of Figure~\ref{fig:timing}. A close inspection suggests a further modulation of the residuals on timescales longer than the binary orbital period. We tested this hypothesis by comparing a constant against a constant plus a sinusoidal function. The latter showed an F-test probability of $\sim1.7\times 10^{-3}$, corresponding to a $\sim3\sigma$ statistical improvement due to the additional component characterised by an amplitude of $(3.3\pm0.8)\times 10^{-2}$ phase cycles, and period $1.99\pm0.14$ days.   

\begin{figure}
  \includegraphics[width=0.4\textwidth]{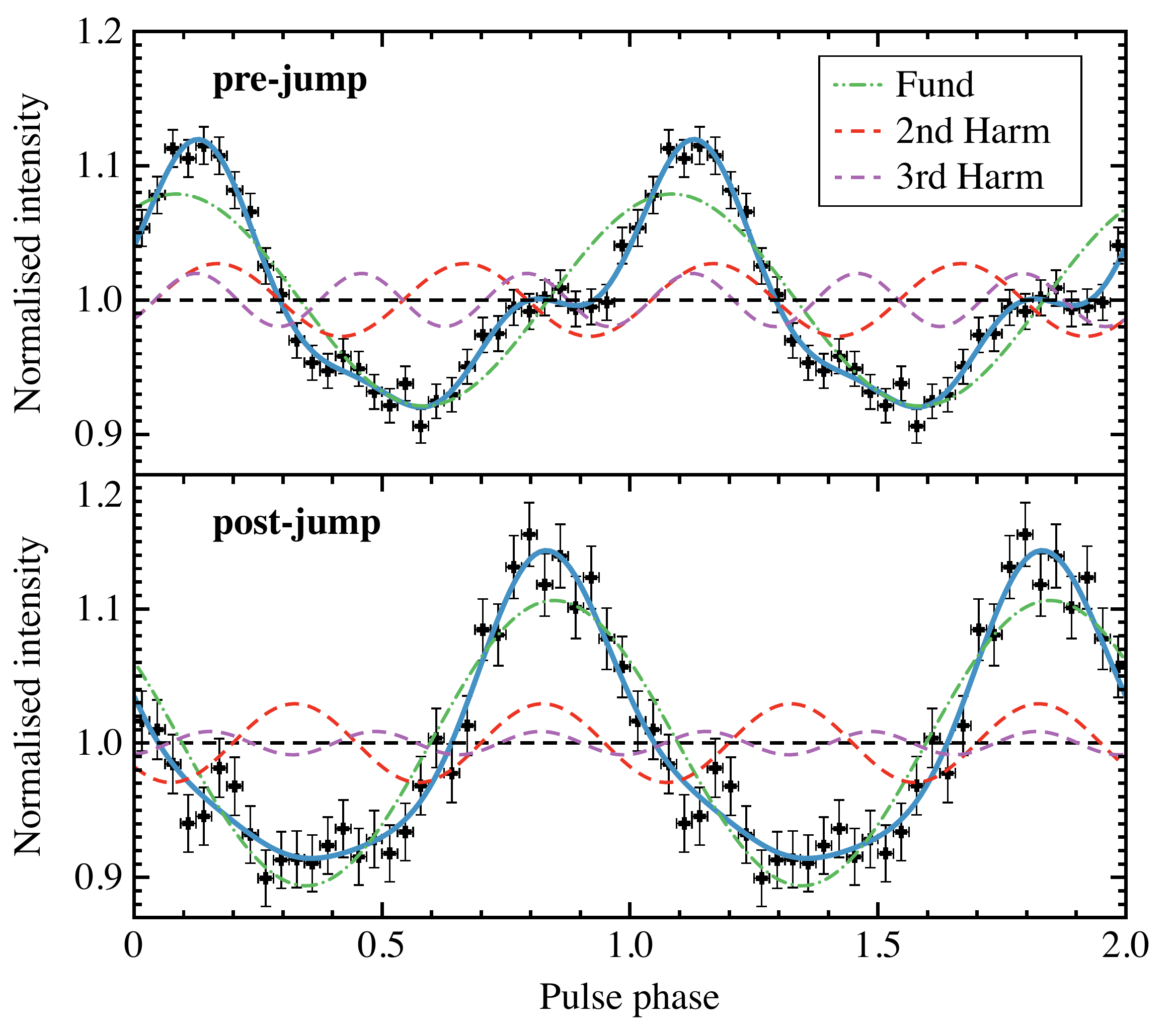}
  \caption{Average pulse profiles (black points) generated combining the \nicer{} data pre- (top-panel) and post- (bottom-panel) appearance of the pulse phase jump (MJD 59750.2) after correcting for the best-fit orbital parameters reported in Table~\ref{tab:solution}. The best-fitting model (cyan solid line) is well described by the superposition of three harmonically related sinusoidal functions. Two cycles of the pulse profile are shown for clarity.}     
  \label{fig:profile}
  \end{figure}

To further investigate the effect of the phase jump, we generated the average pulse profile pre and post-phase jump (Figure~\ref{fig:profile}). Both profiles are well described as the superposition of three harmonically related sinusoidal functions. The fundamental, second, and third harmonics of the pre-jump profile are characterised by fractional amplitudes of $(7.9\pm0.2)$\%, $(2.7\pm0.2)$\%, and $(2.0\pm0.3)$\%, respectively. The post-jump profile presents fractional amplitudes of $(10.6\pm0.4)$\%, $(2.9\pm 0.5)$\%, and $(0.9\pm0.4)$\% for the fundamental, second and third harmonics, respectively.

In the third panel of Figure~\ref{fig:timing}, we report the evolution of the background-corrected fractional amplitude of the pulse profile estimated from the best-fitting solution (filled circles), as well as upper limits for non-detection (filled triangles). The fractional amplitude remains almost constant around the value of $7.5\%$ for the first two days when it starts to increase, reaching a value of $\sim32\%$ during the last signal detection. 

\begin{table}
\caption{Orbital parameters and spin frequency of \maxi{} with uncertainties on the last digit quoted at 1$\sigma$ confidence level. $T_0$ represents the reference epoch for this timing solution.}
\centering
\begin{tabular}{l c}
Parameters  &  \\
\hline
\hline
R.A. (J2000) & $19^h56^m39.11^s \pm 0.04^s$\\
Decl. (J2000) & $03^\circ26' 43.7\arcsec \pm 0.6\arcsec$\\
$P_\mathrm{orb}$ (s) &3653.046(61)\\
$x$ (lt-s) &0.013796(25)\\
$T_{ASC}$ (MJD/TDB) & 59749.633146(18)\\
Eccentricity &$ < 1.4 \times 10^{-2}$ (3$\sigma$~c.l.)\\
$\nu_0$ (Hz) &313.64374049(22)\\
$T_0$ (MJD/TDB) & 59749.0\\
\hline
$\chi^2_\mathrm{red}$/d.o.f. & 1.23/23\\
\end{tabular}
\label{tab:solution}
\end{table}

\subsection{Spectral analysis}
\label{sec:spectra}
We investigated the spectral properties of \maxi{} by generating four energy spectra along the decaying phase of the outburst. Data intervals selected to create the spectra are shown in Figure~\ref{fig:timing}. To perform the spectral analysis, we set \citet{Wilms00} elemental abundances and \citet{Verner96} photo-electric cross-sections.   
The 0.5--10 keV energy spectra are well modelled by an absorbed thermal component (black-body) combined with a power-law continuum (\texttt{TBabs*[bbodyrad+powerlaw]} in Xspec).

The best-fit values of N$_{\rm H}$ vary from $(0.9\pm0.1)\times 10^{21}$ cm$^{-2}$ (obtained in \#1) up to an average value of $(2.5\pm0.5)\times 10^{21}$ cm$^{-2}$ for the other intervals. 
We noticed that the Galactic absorption in the direction of the source is estimated to be $\sim1 \times 10^{21}$ cm$^{-2}$ \citep{HI4PI-Collaboration:2016ut}. Moving from \#1 to \#4, the black-body temperature (kT$_{BB}$) decreased significantly from $(0.45\pm0.01)$~keV down to $(0.26\pm0.03)$~keV, as well as its normalisation that varies from $(5.3\pm2.2)$~km to $1.6^{+2}_{-1}$~km \citep[estimated at $5\pm2$~kpc;][]{Ravi:2017tl}, compatible with a fraction of the NS surface. The power-law photon index at the peak is $\Gamma=1.58\pm0.04$, compatible with the source being in a hard state. As expected, during the descending phase, $\Gamma$ increases, reaching the final value of $2.81\pm0.03$. We detected clear correlations between N$_{\rm H}$, kT$_{BB}$ and $\Gamma$ by Goodman-Weare algorithm of Monte Carlo Markov Chain \citep{Goodman2010} characterised by 20 walkers and chain length of $10^6$. However, the observed spectral evolution cannot be explained only by model degeneracy.
We found no evidence for spectral lines (e.g., iron K-$\alpha$) nor reflection features. Finally, the unabsorbed 0.5--10~keV flux decreases from $(2.95\pm0.02)\times 10^{-10}$~erg~cm$^{-2}$~s$^{-1}$ to $(8.06\pm0.08)\times 10^{-12}$~erg~cm$^{-2}$~s$^{-1}$ close to the quiescence phase. 

\begin{figure}
	 \resizebox{0.45\textwidth}{!}{
  \includegraphics[angle=0]{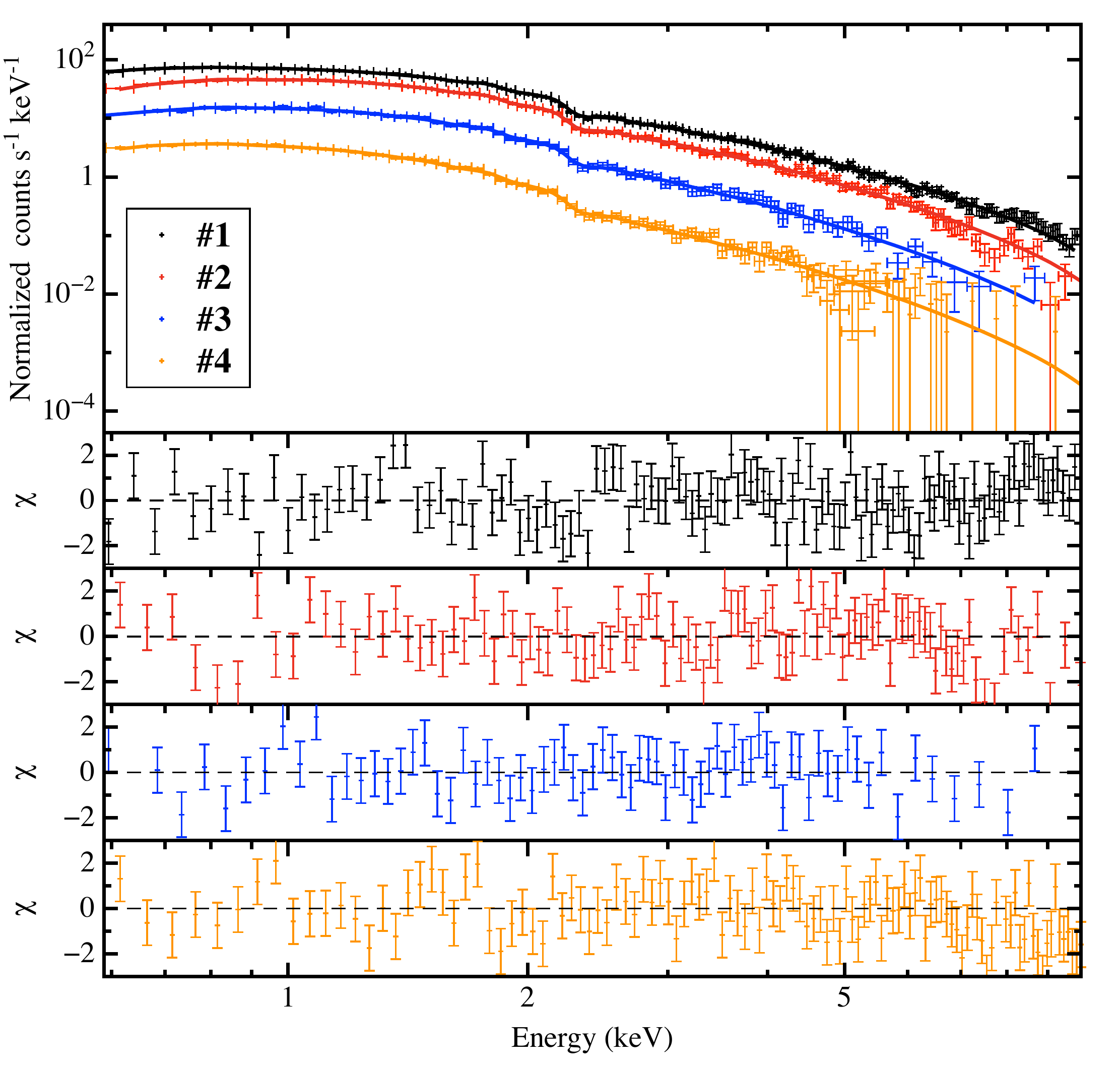}
  }
  \caption{\emph{Upper panel}: \nicer{} spectra of \maxi{} obtained by selecting four intervals (\#1--\#4) during the descending phase of its latest outburst. Black, red, blue and orange represent intervals \#1, \#2, \#3 and \#4, respectively. Solid lines show the best-fitting model to the data. \emph{Lower panels}: residuals with respect to the best-fit models expressed in units of standard deviations for each of the spectra analysed.}
  \label{fig:spectrum}
\end{figure}

\section{Discussion}
\label{sec:discussion}
We reported on the temporal and spectral properties of the newly discovered AMXP \maxi{} during its latest outburst as monitored by \nicer{}. The detection of coherent X-ray pulsations at $\sim314$~Hz allowed us to identify the accreting compact object as a rapidly-rotating NS, confirming the hypothesis drawn from the previous outbursts \citep[see, e.g.,][]{Mata-Sanchez:2017vl,Ravi:2017tl,Beri:2019va}. We interpreted the sinusoidal drift of the X-ray pulsation as the result of the Doppler modulation of the pulsation frequency in a binary system with an orbital period of $\sim1$ hour. Phase-coherent timing analysis allowed us to refine the orbital ephemeris (Table~\ref{tab:solution}). 

Pulse phase delays estimated from the best-fit timing solution show a clear jump of $\sim$0.2 phase cycles. Similar shifts have been reported for SAX J1808.4$-$3658 \citep{Burderi:2006va}, XTE J1814$-$338 \citep{Papitto:2007wp}, XTE J1807$-$294 \citep{Riggio:2008wz,Patruno:2010wi}, SWIFT J1749.4$-$2807 \citep{Sanna:2022tt}, and MAXI J1816$-$195 (Bult et al. 2022, submitted). As shown in Figure~\ref{fig:profile}, the average pulse profile loses harmonic content after the phase jump, with an increase in the fractional amplitude of the fundamental component and a decrease of the third harmonic (detected at a 2$\sigma$ c.l.). No correlation between the phase jump and X-ray count rate seems to exist. The origin of phase jumps in AMXPs is still an open question, and the investigation of the mechanisms proposed to explain them is beyond the scope of this work \citep[see, e.g.,][for some of the proposed scenarios]{Bildsten:1998wb, Lamb:2009wd, Poutanen:2009wb,Riggio:2011ua, Long2012}. Moreover, we found marginal evidence of modulation in the phase residuals with a $\sim2$ days period and amplitude of $\sim0.03$ phase cycles. If confirmed, this could signify a planet-like object (with mass $\lesssim10^{-3}$~M$_{\odot}$) gravitationally bound to the binary system.  

We attempted to set a preliminary constraint on the NS dipolar magnetic field. Assuming the spin equilibrium for the X-ray pulsar, we can then express the magnetic field as:
\begin{equation}
\label{eq:spineq}
B=0.63\,\zeta^{-7/6}\left(\frac{P_{\text{spin}}}{2\text{ms}}\right)^{7/6}\left(\frac{M}{1.4M_{\odot}}\right)^{1/3}\left(\frac{\dot{M}}{10^{-10}M_{\odot}/\text{yr}}\right)^{1/2}10^8~\text{G},
\end{equation}
where $\zeta$ (generally between 0.1--1) corresponds to the ratio between the magnetospheric radius and the Alfv\'en radius \citep[see, e.g.,][]{Ghosh79a,Wang96}, $P_{\text{spin}}$ and $M$ are the pulsar spin period and mass, respectively, and $\dot{M}$ represents the mass accretion rate onto the NS. Considering the peak unabsorbed flux (0.5--10~keV) and a source distance of $5\pm2$~kpc \citep{Ravi:2017tl}, we infer $\dot{M}\simeq0.76\times10^{-10}$ M$_{\odot}$~yr$^{-1}$ for a NS radius and mass of 1.4~M$_{\odot}$ and 10~km, respectively. 
From Eq.~\ref{eq:spineq}, we estimate a dipolar magnetic field ranging between $1\times10^8$~G and $1.4\times 10^{9}$~G, in line with the typical magnetic field of known AMXPs \citep[see, e.g.,][]{Mukherjee:2015td}.

The NS mass function $f(m_2, m_1, i)\sim1.6 \times 10^{-6}$~M$_{\odot}$, combined with the absence of total eclipses or dips in the X-ray light curve \citep[inclination $\lesssim 60^{\circ}$; see, e.g.,][]{Frank1987} suggest a lower limit on the companion star mass of $m_2 \gtrsim 1.7 \times 10^{-2}$~M$_{\odot}$ (for a 1.4~M$_{\odot}$ NS), which increases up to $m_2 \gtrsim 2.1 \times 10^{-2}$~M$_{\odot}$ if we consider a 2~M$_{\odot}$ NS.

\begin{figure}
  \includegraphics[width=0.4\textwidth]{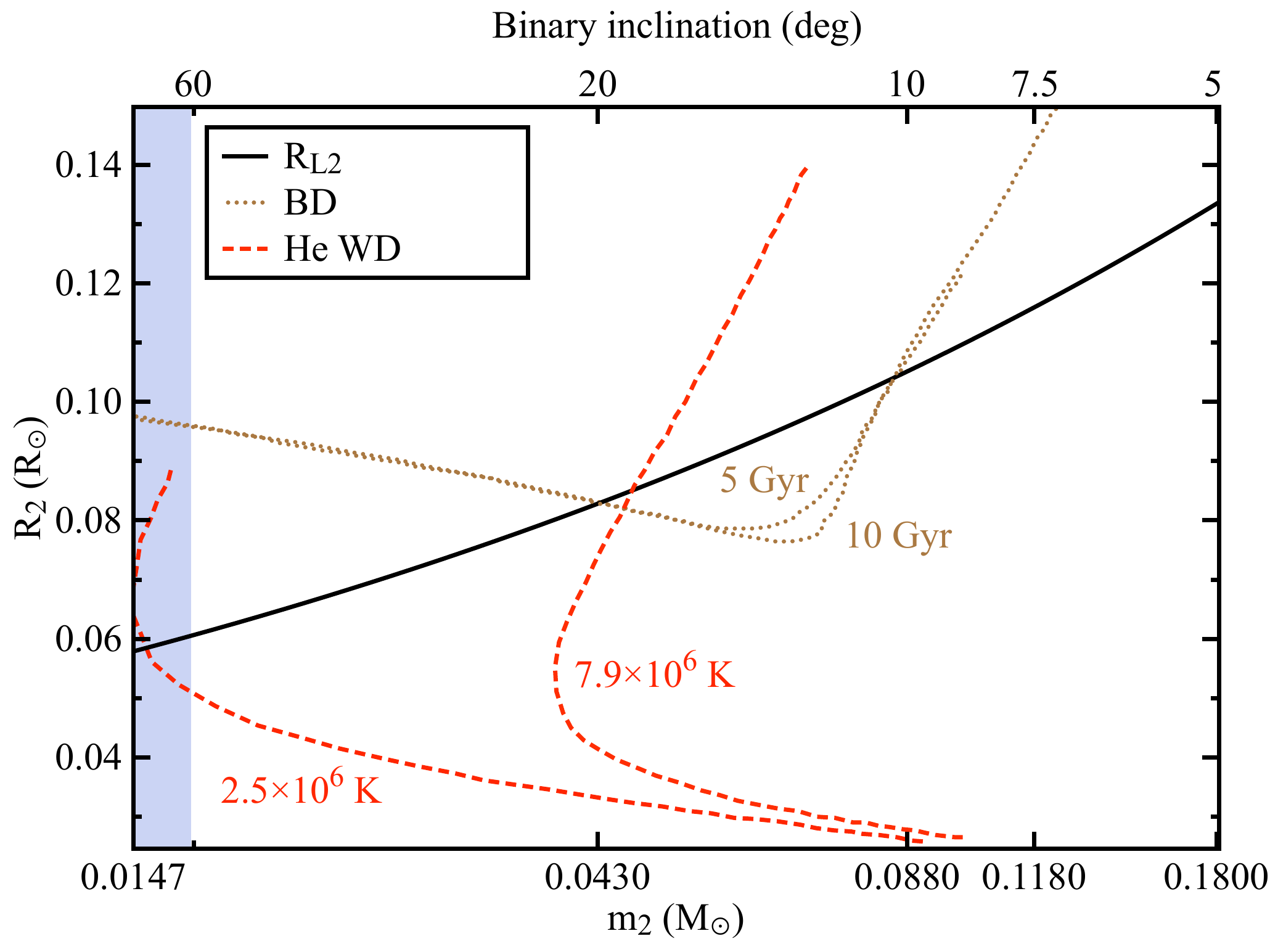}
  \caption{Radius-mass plane showing the size constraints on the Roche-lobe-filling companion star \maxi{} (black solid line). The blue area defines the mass constraints for inclination angles between 60 and 90 degrees. Red-dashed lines represent theoretical mass-radius relations for hot ($7.9\times 10^6$ K) and warm ($2.5\times 10^6$ K) Helium white dwarfs. Brown-dotted lines show low-mass main sequence/brown dwarfs of age 5 and 10 Gyr for solar metallicity abundances. The top axis indicates the corresponding binary inclination angle in degrees for a 1.4 M$_\odot$ NS.}     
  \label{fig:mass}
\end{figure}

We derive the donor radius as a function of its mass, $R_2\simeq 0.2\,m_2^{1/3}\,P_{orb,1h}^{2/3}$ R$_{\odot}$, by combining the Roche-Lobe overflow contact condition ($R_2\approx R_{L2}$) with the NS mass function. To further investigate the nature of the donor star, in Figure~\ref{fig:mass}, we compare the latter expression (solid-black line) with different types of low-mass stars. Red-dashed lines represent theoretical mass-radius relations for warm ($2.5 \times 10^6$~K) and hot ($7.9\times 10^6$~K) Helium white dwarfs \citep[He WD;][]{Deloye2003a}. Warm He WDs seem unlikely given the required inclination $\sim 90^{\circ}$. On the other hand, hot He WDs would imply a donor mass $\sim0.045$ M$_{\odot}$, with an orbital inclination of $\sim20^{\circ}$. Brown-dotted lines represent numerically simulated mass-radius relations for brown dwarfs at 5 and 10 Gyr \citep{Chabrier:2009vh}. Intersections between the curves suggest a brown dwarf donor star with mass of 0.043$-$0.085~M$_{\odot}$, which corresponds to an orbital inclination between 20 and 10 degrees. However, we cannot exclude the possibility of the donor being bloated with respect to its thermal equilibrium radius because of irradiation from the compact object. Further studies are needed to favour one of the proposed scenarios. Nevertheless, these constraints on the donor mass strongly support the scenario for which the late-K/early-M dwarf star identified as the optical counterpart of \maxi{} is not the donor star of the detected AMXP, but is instead a member of a triple system with an ultracompact AMXP binary \citep{Ravi:2017tl}. 


Finally, the energy spectrum is well described by an absorbed soft black-body-like component ($kT\sim 0.4$ keV) compatible with emission from the NS surface plus a power-law-like component characterised by $\Gamma \sim 1.6$ in line with typical thermal Comptonised components observed in the hard state of AMXPs \citep[see, e.g.][]{Falanga05a,Gierlinski2005a,Sanna:2017tx}. The cooling of the black-body temperature and the increase of the power-law photon index as the source X-ray activity dims out are in line with the spectral evolution of the previous outbursts of the source \citep{Beri:2019va}, as well as other AMXPs \citep[see, e.g., ][]{Ng:2021wy, Sanna:2018wh}. We found no evidence for iron K-$\alpha$, nor reflection components in the \nicer{} band. Interestingly, a similar result has been reported for at least other four \emph{ultra-compact} AMXPs, i.e., IGR J16597$-$3704 \citep{Sanna:2018td}, XTE J1807$-$294 \citep{Campana:2003wb}, XTE J1751$-$305 \citep{Miller03}, and SWIFT J1756.9$-$2508 \citep[see, e.g.,][]{Sanna:2018aa,Koliopanos:2021wl}. However, evidence for iron lines has been reported for compact systems such as NGC 6440 X$-2$ \citep{Heinke:2010tt}, MAXI J0911$-$655 \citep{Sanna:2017ul}, and IGR J17062$-$6143 \citep{Degenaar:2017ts}. 
Assuming a source distance $d=5\pm2$~kpc \citep{Ravi:2017tl}, we constrain the peak luminosity during the latest outburst in the range $L=(3.2-17.2)\times 10^{35}$~erg~s$^{-1}$ taking into account the distance uncertainty, consistent with \maxi{} being a very faint X-ray transient \citep[see, e.g.,][]{Wijnands06,Beri:2019va}.

\section*{Acknowledgements}
PB acknowledges support from NASA through the NICER Guest Observer Program and the CRESST II cooperative agreement (80GSFC21M0002). NICER work at NRL is supported by NASA.

\section*{Data availability }
The data utilised in this article are publicly available at \href{https://heasarc.gsfc.nasa.gov/cgi-bin/W3Browse/w3browse.pl}{https://heasarc.gsfc.nasa.gov/cgi-bin/W3Browse/w3browse.pl}, while the analysis products will be shared on reasonable request to the corresponding author.


\bibliographystyle{mnras}
\bibliography{biblio.bib}







\bsp	
\label{lastpage}
\end{document}